\documentclass[aps,prl,twocolumn,superscriptaddress,letterpaper]{revtex4-2}

\usepackage{graphicx}
\usepackage{bm}
\usepackage{amssymb}
\usepackage{color}
\usepackage{amsmath}
\usepackage{ulem}
\usepackage[colorlinks=true,linkcolor=blue,anchorcolor=black,citecolor=blue,filecolor=black,menucolor=black,runcolor=black,urlcolor=blue]{hyperref}

\begin{document}

\title{Momentum-Space Mismatch Enables Non-Hermitian Valley Filtering}

\author{Kai Shao}
\email{shaokai@ynu.edu.cn}
\affiliation{School of Physics and Astronomy, Yunnan Key Laboratory for Quantum Information, Yunnan University, Kunming 650091, People’s Republic of China}
\affiliation{National Laboratory of Solid State Microstructures, School of Physics, Jiangsu Physical Science Research Center, and Collaborative Innovation Center of Advanced Microstructures, Nanjing University, Nanjing 210093, China}

\begin{abstract}
Valley-selective transport is central to valleytronics, but existing valley filters are largely tailored to specific materials or lattice geometries, motivating the search for a common microscopic ingredient.
Here we identify the momentum-space mismatch between the band-touching nodes of a system and its dissipative reservoir as a route to node-selective non-Hermitian skin effects (NHSE).
When the two are coupled, the mismatch is encoded in the
retarded self-energy, producing node-dependent dissipation and
contrasting quasiparticle lifetimes. 
The resulting effective non-Hermitian Hamiltonian exhibits 
opposite point-gap spectral winding at the two nodes, 
giving rise to a node-resolved NHSE with strongly asymmetric dissipation rates: 
one node decays rapidly into the reservoir 
while the other survives over the device, yielding a nearly valley-polarized outgoing current. 
We demonstrate this mechanism in a minimal one-dimensional
dimer chain and in two-dimensional systems hosting genuine Dirac
valleys, including $\pi$-flux rectangular and honeycomb lattices.
In all cases, transport calculations yield nearly 100\% valley
polarization, while remaining gate-tunable
and robust against disorder and edge imperfections. 
Our results show that momentum-space mismatch provides a route to non-Hermitian valley filtering.
\end{abstract}
\maketitle

\textit{Introduction.}---Valleytronics~\cite{Beenakker2007NP,xu2014spin,
schaibley2016NRM,vitale2018small} exploits the valley degree of freedom 
as an information carrier, offering a compelling analog to 
spintronics~\cite{DSarma2004RMP}. In graphene and related hexagonal 
materials, the $K$ and $K'$ valleys are well-separated in reciprocal 
space, enabling a rich variety of valley-resolved phenomena, including 
the valley Hall effect~\cite{NIu2007PRL,yamamoto2015PSJ,gorbachev2014Science,
mak2014Science}, optical valley polarization~\cite{mak2012NaNo,zeng2012NaNo,
cao2012Nc,jones2013NaNo}, and the valley Zeeman 
effect~\cite{srivastava2015NP,aivazian2015NP,LiYilei2014PRL,
MacNeill2015PRL,stier2016Nc}. Central to any practical valleytronic 
application is a robust valley filter capable of generating fully 
valley-polarized currents.

Various proposals for valley-polarized currents have been put forward, including zigzag ribbon modes~\cite{Beenakker2007NP}, line defects~\cite{GunlyckePRL2011,LiuYangPRB_2013,ChenJHPRB2014}, zero-line modes~\cite{Martin08prl,QiaoZhenhuaPRL2014,qiao2011NanoLetter,li2018valley}, strain engineering~\cite{fujita2010AIP,JiangPRL2013,zhai2011NJP}, and nanobubbles~\cite{Settnes16prl}. These approaches exploit distinct mechanisms, such as boundary states, interface modes, and strain-induced pseudomagnetic fields, and their performance can depend on the device geometry, carrier injection conditions, and susceptibility to intervalley scattering. Achieving simultaneously efficient and robust valley filtering therefore remains an important challenge~\cite{schaibley2016NRM}.
Recent advances in non-Hermitian physics have revealed the non-Hermitian skin effect (NHSE)~\cite{YaoShunyu2018PRL,YaoShunyu2018PRL2,Kunst2018PRL,Yokomizo2019PRL,YangZhesen2020PRL,Shao_2022_PRB,Genghao_2023PRB}, whereby the point-gap topology of a complex spectrum~\cite{Genghao_2023PRB,Borgnia2020PRL,Okuma2020PRL,ZhangKai2020PRL,Kunst_RMP_2021} can drive the accumulation of bulk eigenstates at open boundaries and give rise to transport phenomena without Hermitian counterparts. These developments suggest a route to valley filtering based on valley-dependent non-Hermitian transport~\cite{Shao_2024_PRL_Moire}. In Ref.~\cite{Shao_2024_PRL_Moire}, valley-selective transport associated with the NHSE was demonstrated in heterostrained bilayer graphene, where strain-induced shifts of the Dirac cones and moir\'e Umklapp coupling generate valley-dependent dissipation. This raises an open question: are moir\'e Umklapp processes and strain-induced valley-dependent dissipation essential for non-Hermitian valley filtering, or can the effect arise from a simpler ingredient? Here we show that a momentum mismatch between the band-touching nodes of a system and a dissipative reservoir can produce node-dependent dissipation and valley filtering.

\begin{figure}
	\centering
	\includegraphics[width=0.5\textwidth]{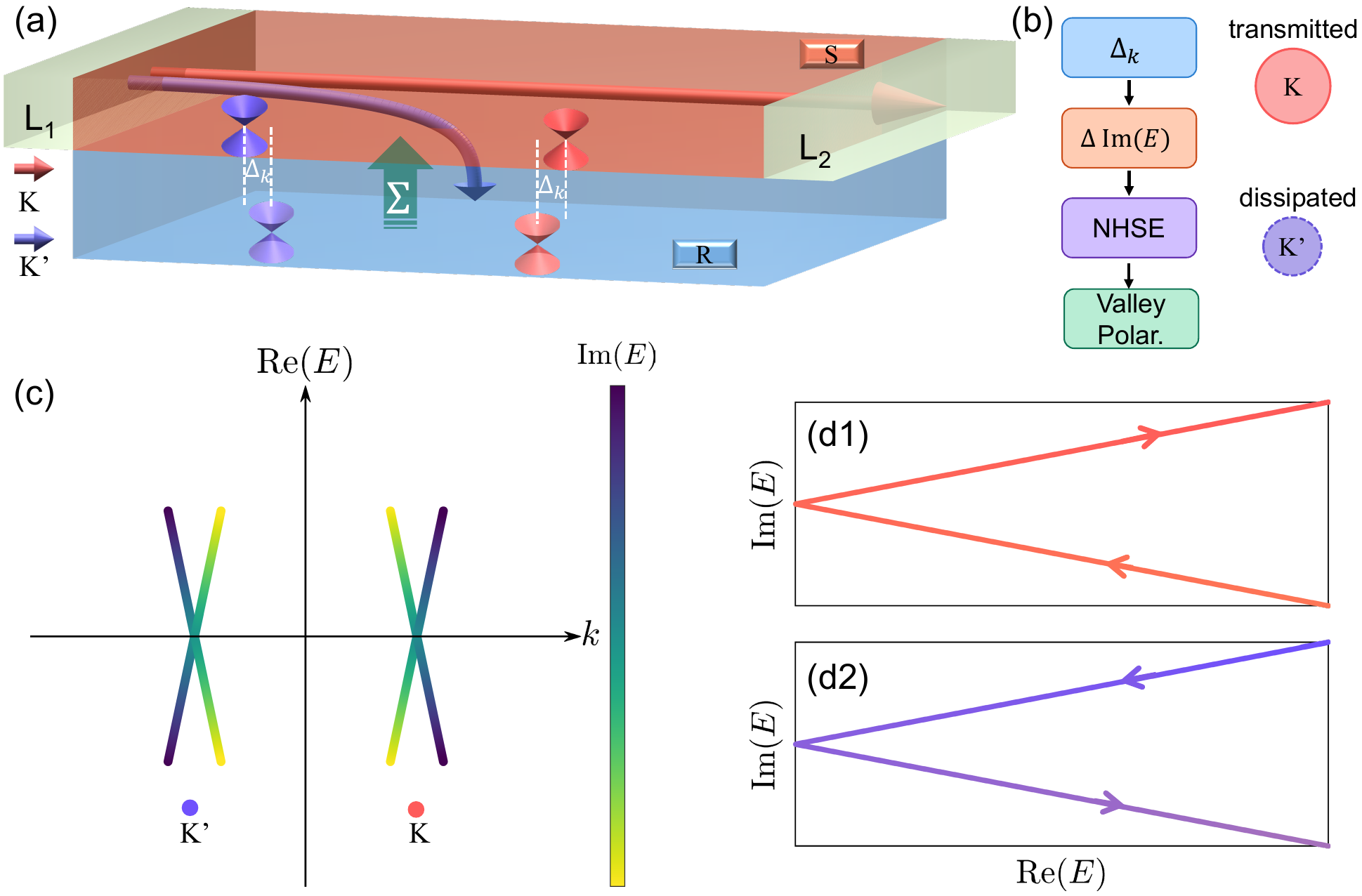}
	\caption{(a) Schematic of a system (S) with momentum-separated
        band-touching nodes coupled to a reservoir (R). The momentum
        mismatch $\Delta_k$ maps onto a node-dependent self-energy
        $\Sigma$, enabling selective transmission of one node while
        filtering the other. (b) Node-selective non-Hermitian response:
        $\Delta_k$ induces asymmetric imaginary self-energies
        $\mathrm{Im}(\Sigma)$, producing node-contrasting quasiparticle
        lifetimes and triggering the NHSE---transmitting $K$ electrons
        while dissipating $K'$ electrons into the reservoir.
        (c) Effective complex band structure of S: $\mathrm{Re}(E)$
        retains the Dirac dispersion, while $\mathrm{Im}(E)$ shows
        node-dependent dissipation. (d1,d2) Complex-energy spectra for
        $K$ and $K'$ nodes. Opposite spectral winding directions dictate
        the node-contrasting NHSE and directional transport asymmetry.
}
	\label{fig1}
\end{figure}

In this Letter, we identify such a mechanism based on a single physical 
ingredient: a momentum-space offset $\Delta_k$ between the band-touching 
nodes of a system $S$ and its dissipative reservoir $R$ [Fig.~\ref{fig1}(a)]. 
When $S$ is coupled to $R$, this offset is encoded in the retarded 
self-energy $\Sigma$, which becomes strongly node-asymmetric 
[Fig.~\ref{fig1}(b)]: the two momentum-space nodes acquire contrasting 
imaginary self-energies, corresponding to node-contrasting quasiparticle 
lifetimes. The resulting effective non-Hermitian Hamiltonian features 
complex spectra that wind in opposite directions for the two nodes 
[Figs.~\ref{fig1}(d1,d2)]---the hallmark signature of a node-resolved 
NHSE [Fig.~\ref{fig1}(c), SM~\cite{SM}]. Physically, electrons injected from one 
electrode dissipate into the reservoir at markedly different rates 
depending on node: carriers at the strongly dissipative node are 
lost before reaching the opposite electrode, while those at the 
weakly dissipative node largely survive the transit, yielding a 
near-perfectly polarized outgoing current.

We demonstrate this in three lattice models of increasing
physical richness. A one-dimensional dimer chain serves as the
minimal realization of the $\Delta_k$ mechanism, where the role of
momentum-space nodes is played by band-touching points controlled
by hopping parameters. We then extend the mechanism to two-dimensional
systems hosting Dirac valleys: a $\pi$-flux rectangular
lattice and a honeycomb lattice. In all cases, quantum transport
calculations yield nearly 100\% valley polarization over wide
parameter ranges, with robustness against Anderson disorder and
edge imperfections rooted in bulk non-Hermitian topology.

\begin{figure}
	\centering
	\includegraphics[width=0.5\textwidth]{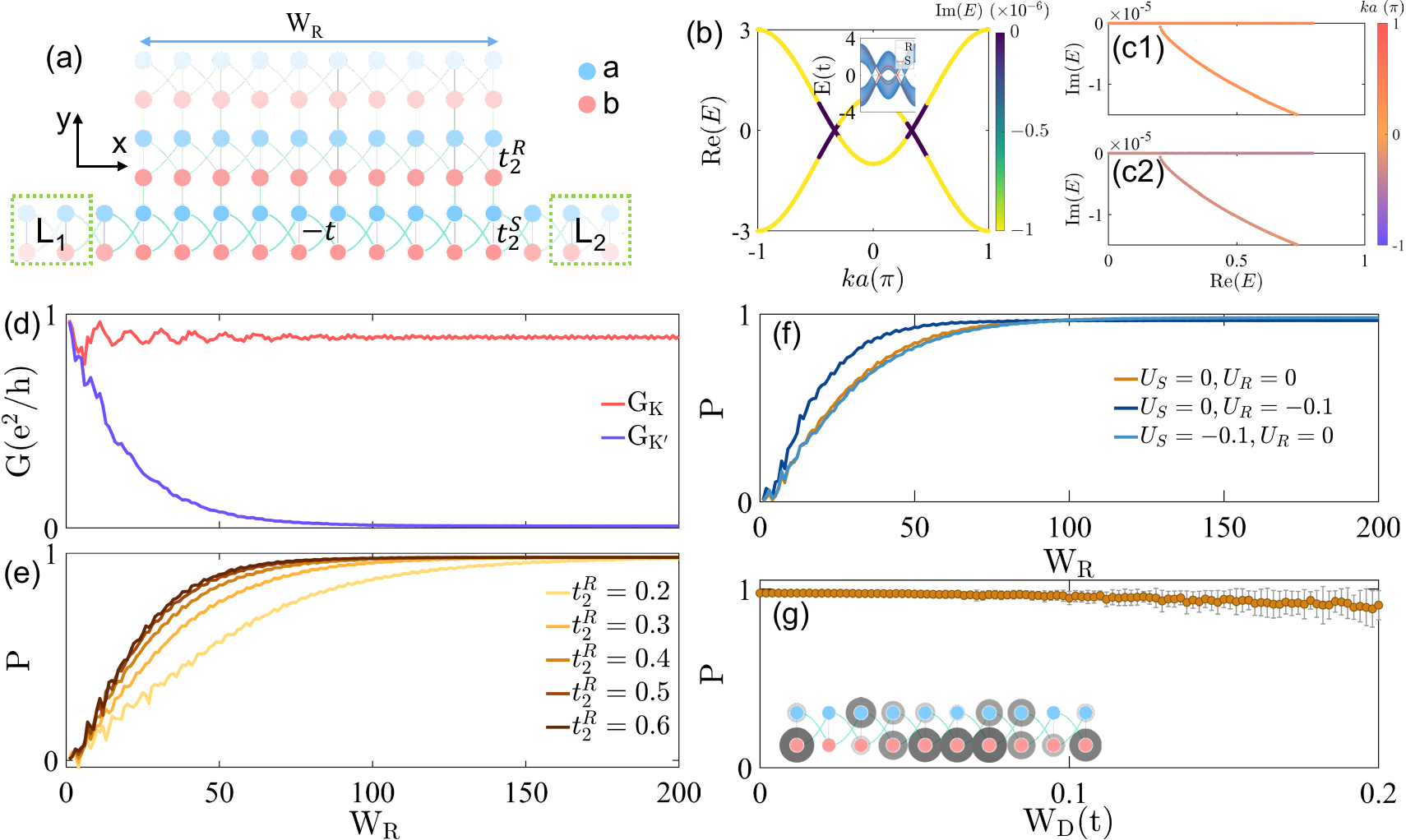}
     \caption{(a) Schematic of the 1D system chain $S$ coupled to an electronic reservoir $R$ (width $W_R$). The intracell hopping amplitudes in $S$ and $R$ are denoted by $t_2^S$ and $t_2^R$, respectively. The system is connected to two leads ($L_1$, $L_2$) for quantum transport measurements. (b) Effective complex energy spectrum of the system chain, with the coupling between $S$ and $R$ scaled to $1\%$ of its actual value for visual clarity. The inset shows the decoupled band structures of $S$ and $R$. (c1), (c2) Zoom-in views of the complex energy spectra in the $\text{Re}(E)\text{-}\text{Im}(E)$ plane near the two nodes, where the color scale indicates the momentum $ka$. (d) Node-resolved differential conductance $G_K$ (red) and $G_{K'}$ (purple) as a function of the reservoir width $W_R$. (e) Node polarization $P$ as a function of $W_R$ for different reservoir hopping strengths $t_2^R$. (f) Tuning of the node polarization $P$ by the gate voltages $U_S$ and $U_R$. (g) Robustness of the node polarization $P$ against Anderson disorder in $S$, as a function of the disorder strength $W_D$ with $W_R=128a$. The relevant parameters are bias voltage $eV=0.36t$, $t_2^S=t$ and $t_2^R=0.4t$.}
	\label{fig2}
\end{figure}

\textit{Cross-coupled dimer chain.}---We first illustrate the mechanism 
in its simplest realization: a one-dimensional dimer chain $S$ coupled 
laterally to an electronic reservoir $R$, with two Leads attached for 
transport measurements [Fig.~\ref{fig2}(a)]. The isolated $S$ and $R$ 
are each described by the Bloch Hamiltonian
\begin{equation}\label{Hk_Chain}
\mathcal{H}_1^{S,R}(k)=\mathbf{d}^{S,R}(k)\cdot\boldsymbol{\sigma},
\end{equation}
where $\boldsymbol{\sigma}$ denotes the Pauli matrices and 
$\mathbf{d}^{S,R}(k)=(t_2^{S,R}-2t\cos ka,\,0,\,0)$, with $-t$ the 
cross hopping between neighboring unit cells and $t_2^{S,R}$ the 
intracell hopping along the vertical direction. 
The dispersion 
$E^{S,R}(k)=\pm(t_2^{S,R}-2t\cos ka)$ hosts band-touching points satisfying 
$2t\cos k_D a = t_2^{S,R}$; we set $t=1$ and $a=1$ throughout. 
The band-touching point position $k_D$ is directly controlled by the ratio 
$t_2^{S,R}/2t$, so a finite mismatch $t_2^S \neq t_2^R$ introduces a 
controllable momentum offset $\Delta_k$ between the band-touching points of 
$S$ and $R$ in reciprocal space, as shown in the inset of 
Fig.~\ref{fig2}(b).

Upon coupling, electrons in $S$ can leak into $R$, generating an 
effective dissipation that fundamentally modifies the spectral and 
transport properties of $S$~\cite{Genghao_2023PRB,Shao_2024_PRL_Moire}. 
This is captured by the retarded Green's function
\begin{equation}\label{Hamiltonian_Eff}
g_S^r(\omega)=\left(\omega-H_S^{\rm{eff}}\right)^{-1}, \quad
H_S^{\rm{eff}}(\omega)=H_S+\tilde{\Sigma}^r_R(\omega),
\end{equation}
where the non-Hermitian self-energy 
$\tilde{\Sigma}^r_R(\omega)=\hat{V}{g}_R^r(\omega)\hat{V}^{\dagger}$ 
encodes the reservoir's influence, $\hat{V}$ is the coupling matrix, and 
${g}_R^r(\omega)=(\omega-H_R-\Sigma^r_R)^{-1}$ is the Green's function 
of $R$, with the surface self-energy $\Sigma^r_R$ determined numerically 
via the iterative Green's function method~\cite{SM}.

To analyze the complex band structure, we assume translational invariance 
along $x$ and integrate out the reservoir degrees of freedom following 
Eq.~\eqref{Hamiltonian_Eff}, yielding the momentum-resolved self-energy
$\tilde{\Sigma}^r_R(\omega,k)=\hat{V}[\omega-\tilde{H}_R(k)]^{-1}
\hat{V}^{\dagger}$, with $\tilde{H}_R(k)=H_R(k)+\Sigma^r_R(\omega,k)$. 
The physical complex band structure follows from diagonalizing 
$H_S^{\text{eff}}(\omega,k)=H_S(k)+\tilde{\Sigma}_R^r(\omega,k)$ 
subject to the self-consistency condition $\text{Re}[E(k)]=\omega$. 
The result is shown in Fig.~\ref{fig2}(b): $\text{Re}(E)$ retains the 
linear character of the isolated $S$ near the band-touching 
points, while $\text{Im}(E)$ exhibits a pronounced asymmetry between 
the right- and left-moving branches of each node. This asymmetry originates from the detuning between the states of $S$ and the bands of $R$ at the same momentum, which differs for the two branches of a node, together with their different group velocities~\cite{SM}. The spectral winding of the two nodes is equal in magnitude 
but opposite in sign [Figs.~\ref{fig2}(c1,c2)], directly confirming 
a node-resolved NHSE.

The transport consequence of this node-asymmetric dissipation is 
probed using the scattering-matrix 
approach~\cite{datta1997electronic,SM}. By labeling transport channels 
with the wave vector $\mathbf{k}$ at a given injection energy $\omega$, 
we define the node-resolved differential conductance
\begin{equation}\label{TkTkp}
G_\kappa(eV)=\frac{e^2}{h}\sum_{\mathbf{k}_1}
\sum_{\mathbf{k}_2\in \kappa}|t_{\mathbf{k}_2,\mathbf{k}_1}|^2,
\end{equation}
where $\kappa=K,K'$ labels the node of the transmitted electron 
in Lead 2, with incident electrons from Lead 1 occupying both nodes, 
and $K$ ($K'$) node electrons selected by $k\in[0,\pi/a]$ 
($k\in[-\pi/a,0]$)~\cite{SM}. As shown in Fig.~\ref{fig2}(d), 
$G_{K'}$ decays rapidly with increasing reservoir width $W_R$ while 
$G_K$ remains large, a direct manifestation of the node-asymmetric 
quasiparticle lifetime. The node polarization~\cite{Beenakker2007NP}
\begin{equation}
P=\frac{G_{K}-G_{K'}}{G_{K}+G_{K'}},
\end{equation}
saturates at $P\simeq100\%$ beyond a moderate $W_R$, demonstrating
that the $\Delta_k$ mechanism realizes an efficient node filter.
Figure~\ref{fig2}(e) shows that this behavior persists robustly over a 
wide range of $t_2^R$, confirming that the effect is not fine-tuned to 
a specific reservoir configuration.

The performance of this node filter can be further optimized via gate 
voltage. Applying on-site potentials 
$H_g=\sum_{i\in S}U_S c_i^\dagger c_i + \sum_{i\in R}U_R c_i^\dagger c_i$ 
modifies the relative alignment of the $S$ and $R$ bands, which 
suppresses oscillations in $P$ and substantially extends the parameter 
region of near-perfect polarization, as shown in Fig.~\ref{fig2}(f). 
The oscillations arise from intranode interference between 
the right-moving state and the component reflected at the ends of the overlap 
region. The gate voltages change the band alignment and hence the self-energy 
and the decay rates of the right- and left-moving states; a stronger attenuation of the reflected component 
weakens the interference, while a faster decay of the $K'$ node brings $G_{K'}$ 
to zero at a smaller $W_R$. 
This relaxes the stringent requirement on a specific $W_R$ and makes 
the node filter more amenable to experimental control.

The robustness of the node filter follows directly from the 
non-Hermitian topology underlying the NHSE. Consider an electron injected 
from Lead 1 in the $K$ node: forward inter-node scattering 
converts it to $K'$, which then has a short lifetime and is rapidly 
lost into $R$ before reaching Lead 2; backward inter-node 
scattering returns it evanescently to the source. In neither case 
does the inter-node-scattered component contribute to the outgoing 
current. We verify this quantitatively by introducing Anderson disorder 
$H_{\text{dis}}=\sum_{i\in S}U_i c_i^\dagger c_i$ with 
$U_i\in[-W_D/2,W_D/2]$. As shown in Fig.~\ref{fig2}(g), high node 
polarization is maintained well into the strongly disordered regime, 
confirming the intrinsic resilience of the non-Hermitian node 
filtering mechanism.

\begin{figure*}
	\centering
	\includegraphics[width=1\textwidth]{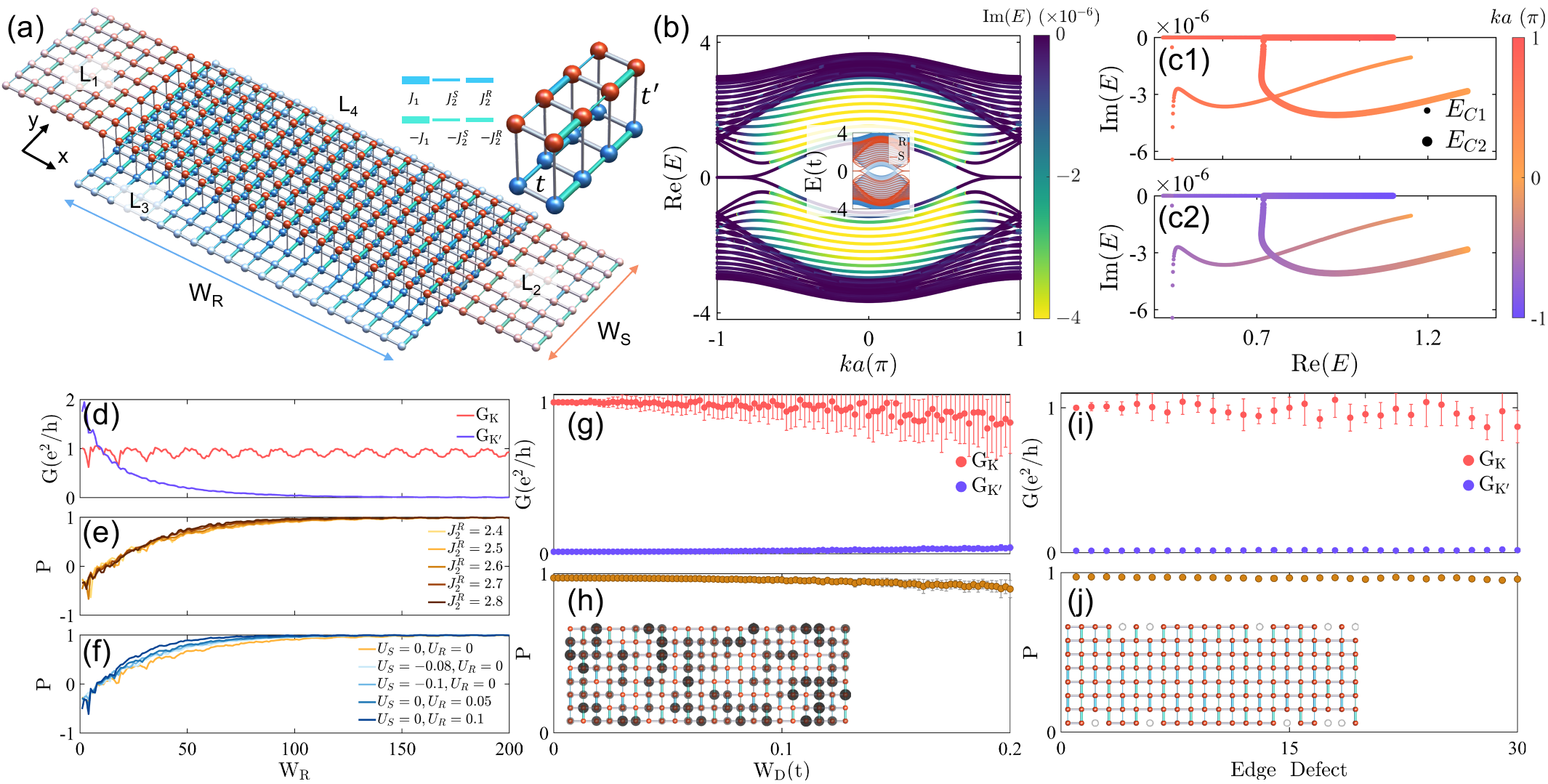}
	\caption{ (color online) Valley-selective transport and robustness of the rectangular lattice heterostructure. 
        (a) Schematic of the interface system comprising an upper layer $S$ and a lower layer $R$, both with rectangular lattices, coupled to four terminals labeled 1--4. Insets show the lattice configuration and hopping amplitudes ($J_1, J_2^S, J_2^R, t', t$) in the overlapping region. 
        (b) Effective complex energy bands of layer $S$ for $W_S = 30$, where the interlayer hopping is scaled to 1\% of its original value for visual clarity. Inset: band structures of uncoupled $S$ and $R$. 
        (c1, c2) Complex energy spectra of the two lowest sub-conduction bands; colors indicate the direction of spectral winding in the complex plane. 
        (d) Valley-resolved differential conductance $G_K(G_{K^{\prime}})$ as a function of $W_R$ at $J_2^R = 2.5$. 
        (e) Valley polarization $P$ vs $W_R$ for various $J_2^R$. 
        (f) Tuning of $P$ via gate voltages $U_S$ and $U_R$. 
        (g--j) Robustness of $G$ and $P$ against (g, h) Anderson disorder and (i, j) edge vacancies for $W_R = 134$. 
        Error bars denote the standard deviation over 24 random realizations. 
        Parameters are $J_1=t, J_2^S=2t, J_2^R=2.5t, t'=0.1t$ and bias voltage $eV = 0.6t$. }
	\label{fig3}
\end{figure*}

\textit{$\pi$-flux rectangular lattice.}---To demonstrate that the 
mechanism extends beyond one dimension, we consider a two-dimensional 
rectangular-lattice heterostructure threaded by a $\pi$ flux per 
plaquette~\cite{Zhaoyuxin_2021_PRL_Gauge_Field,
ZhangBaile_2022_PRL_Projectively}. An upper layer $S$ is coupled to 
a lower reservoir $R$ via interlayer hopping $t'$ [Fig.~\ref{fig3}(a)]. 
Both layers feature uniform hopping $t$ along $x$ and dimerized 
hoppings $J_1$, $J_2^{S,R}$ along $y$, with the $\pi$ flux implemented 
by staggering the signs of the $y$-directional hoppings 
($\pm J_1$, $\pm J_2^{S,R}$). The system is contacted by four terminals 
for transport measurements. The isolated $S$ and $R$ layers share the same form of Bloch 
Hamiltonian~\cite{SM}
\begin{equation}\label{Hk_Square}
\mathcal{H}_2(\mathbf{k}) =
\begin{pmatrix}
0 & J^\dagger & T^\dagger & 0 \\
J & 0 & 0 & T^\dagger \\
T & 0 & 0 & -J e^{i k_y a} \\
0 & T & -J^\dagger e^{-i k_y a} & 0
\end{pmatrix},
\end{equation}
where $J = J_1 + J_2^{S,R} e^{-i k_y a}$ and $T = t(1 + e^{i k_x a})$, 
with $J_2^{S,R}$ taking different values for $S$ and $R$ respectively. 
For $J_1 \neq J_2^{S,R}$, the spectrum 
hosts two Dirac points at $(k_x, k_y) = (\tau\kappa_x, \pi)$, with 
$\tau = \pm 1$ the valley index and momentum $\kappa_x$ determined 
by $2t\cos(\kappa_x a/2) = |J_1 - J_2^{S,R}|$. Setting 
$J_2^S \neq J_2^R$ displaces the Dirac points of $S$ and $R$ from 
each other in momentum space, directly realizing the $\Delta_k$ mismatch 
in two dimensions.

The effective complex band structure of $S$, obtained via the same 
Green's function formalism as in the 1D case with self-energy 
$\bar{\Sigma}_R^{r}(\omega,\mathbf{k})=\hat{V'}[\omega-H_R(\mathbf{k})
-\Sigma^r_3(\omega,\mathbf{k})-\Sigma^r_4(\omega,\mathbf{k})]^{-1}
\hat{V'}^{\dagger}$, is shown in Fig.~\ref{fig3}(b). The imaginary 
part of the energy spectrum again displays pronounced valley asymmetry, 
confirming that the $\Delta_k$ mismatch encodes a valley-dependent 
dissipation rate in two dimensions. The complex spectral winding of the 
two lowest subconduction bands [Figs.~\ref{fig3}(c1,c2)] exhibits 
opposite winding directions for the two valleys, providing the 
topological underpinning of the valley-resolved NHSE.

The transport properties are computed using the same valley-resolved 
formalism as in Eq.~(\ref{TkTkp}). As shown in Fig.~\ref{fig3}(d), 
$G_{K'}$ is strongly suppressed with increasing $W_R$ while $G_K$ 
remains robust, and the valley polarization $P$ approaches unity 
over a broad range of $W_R$ [Fig.~\ref{fig3}(e)]. The behavior 
persists across different values of $J_2^R$, confirming that the 
effect is not sensitive to the precise reservoir parameters. 
As in the 1D case, gate voltages $U_S$ and $U_R$ provide efficient 
tunability: Fig.~\ref{fig3}(f) shows that they suppress oscillations 
in $P$ and expand the regime of near-perfect polarization, making 
the 2D valley filter more robust to parameter variations.

We further examine two types of imperfections relevant to realistic 
2D devices. In the presence of Anderson disorder [Figs.~\ref{fig3}(g,h)], 
both the conductance and valley polarization remain largely unaffected, 
with only moderate fluctuations across 24 disorder realizations (error 
bars). Beyond bulk disorder, we consider random edge vacancies---a 
realistic structural imperfection in 2D devices---and find that $G_K$ 
and $P$ are also nearly unaffected [Figs.~\ref{fig3}(i,j)]. This is 
because valley filtering in our scheme relies on bulk non-Hermitian 
transport rather than fragile edge states: the skin-effect-induced 
preferential absorption operates throughout the bulk of $R$, 
independent of the boundary geometry. The combined resilience against 
both bulk disorder and edge imperfections establishes the practical 
viability of the 2D valley filter.

\begin{figure*}
	\centering
	\includegraphics[width=1\textwidth]{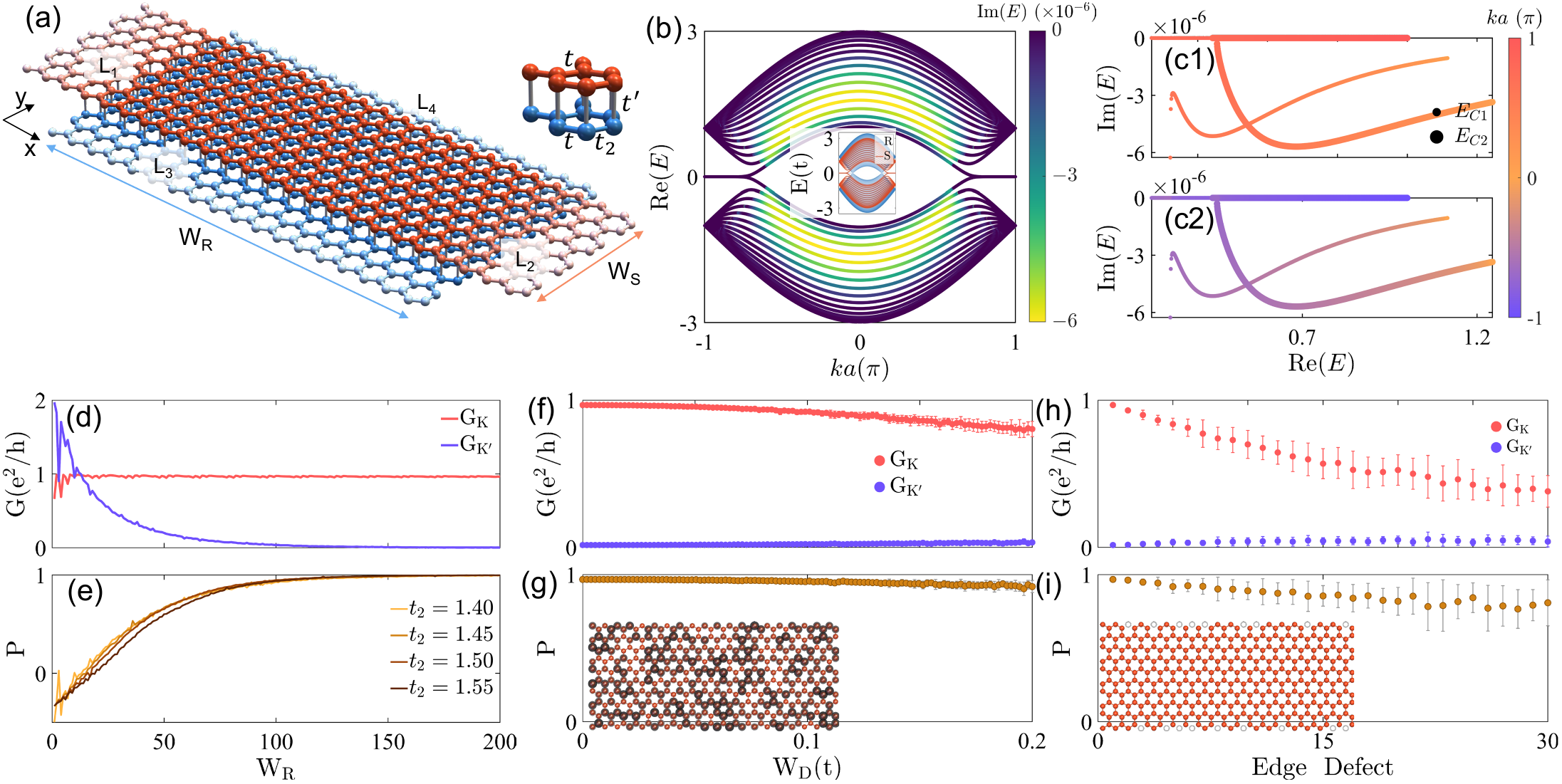}
	\caption{(color online) Valley-selective transport and robustness of the honeycomb heterostructure. 
        (a) Schematic of the proposed interface system consisting of an upper honeycomb layer $S$ and a lower honeycomb layer $R$, coupled to four terminals labeled 1--4. The inset illustrates the interlayer hopping $t'$ and sublattice-dependent intralayer hopping $t_2$ in the overlapping region.
        (b) The effective complex bands of $S$ with the interlayer hopping strength set to 1\% of its original value for better illustration. The inset shows the band structures of the uncoupled $S$ and $R$.  
        (c1,c2) Complex energy spectra of the two lowest subconduction bands (above the bands with zero energy) with the directions of spectral winding marked by the color, respectively.
        (d) Valley-resolved differential conductance $G_K(G_{K^{\prime}})$ as a function of $W_R$ at $t_2 = 1.4t$. 
        (e) Valley polarization $P$ versus $W_R$ for different values of $t_2$.
        Evolution of $G$ and $P$ as a function of Anderson disorder(f), (g) and  number of edge defects(h), (i). Error bars represent the standard deviation over 30 independent realizations. 
        Fixed parameters are $t_2 = 1.4t$,$t' = 0.1t$, $W_R = 119$, $W_S=16$ and bias voltage $eV = 0.4t$.}
	\label{fig4}
\end{figure*}

\textit{Honeycomb lattice.}---Finally, we demonstrate that the 
mechanism applies naturally to systems with intrinsic Dirac valleys 
by considering a honeycomb heterostructure [Fig.~\ref{fig4}(a)]. 
An upper honeycomb layer $S$ is coupled to a lower honeycomb 
reservoir $R$ via interlayer hopping $t'$, with sublattice-dependent 
intralayer hopping $t_2$ in the overlapping region. Four terminals 
are attached for transport measurements. This geometry is directly 
relevant to graphene-based devices, where the $K$ and $K'$ points 
arise intrinsically from the honeycomb structure. The Bloch 
Hamiltonian is~\cite{girvin2019modern}
\begin{equation}\label{Hk_honeycomb}
\mathcal{H}_3(\mathbf{k})=
\begin{pmatrix}
0 & t_2 + t e^{i\mathbf{k}\cdot\mathbf{a}_1} + 
t e^{i\mathbf{k}\cdot\mathbf{a}_2} \\
t_2 + t e^{-i\mathbf{k}\cdot\mathbf{a}_1} + 
t e^{-i\mathbf{k}\cdot\mathbf{a}_2} & 0
\end{pmatrix},
\end{equation}
with primitive vectors $\mathbf{a}_1=(\frac{a}{2},\frac{\sqrt{3}a}{2})$ 
and $\mathbf{a}_2=(-\frac{a}{2},\frac{\sqrt{3}a}{2})$. The two Dirac 
points are located at $(k_x,k_y)=(\tau K_D,0)$, where $\tau=\pm1$ 
labels the valley and $K_D$ is determined by 
$\cos(K_D a)=t_2^2/(2t^2)-1$. Introducing a mismatch in $t_2$ between 
$S$ and $R$ shifts their Dirac points relative to each other in 
momentum space, generating the momentum offset $\Delta_k$ from the 
same physical origin as in the 1D and $\pi$-flux rectangular lattice cases.

Upon coupling, this offset is translated into a valley-dependent 
self-energy $\tilde{\Sigma}_R^r(\omega,\mathbf{k})$, leading to the 
non-Hermitian effective Hamiltonian $H_S^{\rm eff}(\omega,\mathbf{k})$. 
The resulting complex band structure is shown in Fig.~\ref{fig4}(b): 
the imaginary part of the energy displays a clear valley-dependent 
profile, with the two valleys acquiring distinct quasiparticle lifetimes. 
The spectral winding of the two lowest subconduction bands 
[Figs.~\ref{fig4}(c1,c2)] winds in opposite directions for $K$ and 
$K'$, confirming the valley-resolved NHSE in a lattice with intrinsic 
Dirac points. Notably, unlike the 1D case where the momentum-space 
nodes are band-touching points without an intrinsic valley structure, 
here---as in the rectangular-lattice case---the valleys arise from the 
underlying lattice geometry, demonstrating that the $\Delta_k$ mechanism 
applies equally to naturally occurring Dirac materials.

Transport calculations using Eq.~(\ref{TkTkp}) confirm the expected 
filtering behavior. As shown in Fig.~\ref{fig4}(d), $G_{K'}$ is 
rapidly suppressed with increasing $W_R$ while $G_K$ persists, 
producing a strongly valley-polarized current. The valley polarization 
$P$ [Fig.~\ref{fig4}(e)] approaches unity over a wide range of $W_R$ 
and remains robust against variations in $t_2$, confirming that the 
valley filter does not rely on fine-tuning of the hopping parameters.
The robustness of the honeycomb valley filter is examined in 
Figs.~\ref{fig4}(f-i). In the presence of Anderson disorder 
[Figs.~\ref{fig4}(f,g)], the conductance asymmetry between the two 
valleys is well preserved and the polarization $P$ remains close to 
unity, with only moderate fluctuations across 30 independent disorder 
realizations. For edge defects [Figs.~\ref{fig4}(h,i)], even as the number 
of edge vacancies grows, $G_K$ remains largely unaffected and 
$P$ stays near unity. This robustness is expected to improve 
further with increasing $W_S$~\cite{SM}, as the valley filtering 
relies on bulk rather than edge transport. 
These results establish that the 
valley-selective non-Hermitian transport mechanism is fully compatible 
with honeycomb systems and robust against realistic imperfections, 
highlighting its direct applicability to graphene and other 
hexagonal-lattice platforms.

\textit{Conclusions.}---In summary, we have shown that a momentum-space offset $\Delta_k$ between the nodes of a system and its dissipative reservoir provides a route to non-Hermitian valley filtering. It produces node-contrasting quasiparticle lifetimes and opposite spectral winding for the two nodes, i.e.\ a node-resolved NHSE, which yields near-perfect polarization.
We demonstrate this in a one-dimensional dimer chain and in two-dimensional $\pi$-flux rectangular and honeycomb lattices hosting genuine Dirac valleys, without the strain or moir\'e engineering of Ref.~\cite{Shao_2024_PRL_Moire}. The resulting filters reach nearly 100\% polarization over wide parameter ranges, are gate tunable, and are robust against Anderson disorder and edge imperfections.
Experimentally, $\Delta_k$ is set by a hopping mismatch between the system and the reservoir, which can be engineered in topolectrical circuits through capacitor and inductor values, with resistive elements as the reservoir~\cite{Imhof_NP_2018_Topolectricalcircuit,Helbig_2020_NP_GeneralizedBBC,LeeChingHua_2018_Commu_Phy_Topolectrical,teo2024_SciBul_pseudomagnetic}, and in photonic and acoustic lattices through coupling strengths and absorptive elements~\cite{ZhangXiao_2025_PRL_PhotonicFlatLandauLevels,YueSiJie_2026_PRL_UniversalValleyFiltering,DengWeiYing_2026_PRL_ImplementingNonAbelian,Dengweiying_2026_PRL_AcousticBoundPair}. The polarization can be read out from the node-resolved transmission, with an overlap region long compared with the decay length of the filtered node.

\textit{Acknowledgments.}---We thank Wei Chen for helpful discussions. This work was supported by the National Natural Science Foundation of China under Grant No. 12604107 (S.K.), No. 12074172 (W.C.), No. 12222406 (W.C.), Xing Dingyu Academician Workstation of Yunnan Province (202605AF350035), and the National Key Projects for Research and Development of China under Grant No. 2022YFA120470 (W.C.).

\end{document}